# SIMULATION OF MACHINE LEARNING-BASED 6G SYSTEMS IN VIRTUAL WORLDS

Ailton Oliveira[1], Felipe Bastos[1], Isabela Trindade[1], Walter Frazão[1], Arthur Nascimento[1], Diego Gomes[2], Francisco Müller[1] and Aldebaro Klautau[1]

[1]Universidade Federal do Pará - LASSE — www.lasse.ufpa.br, Av. Perimetral S/N, Belém, Pará, Brazil.,
[2]Universidade Federal do Sul e Sudeste do Pará - IGE — www.ige.unifesspa.edu.br , Marabá, Pará, Brazil.

Corresponding author: Ailton Oliveira — Email: ailton.pinto@itec.ufpa.br

***Abstract*** – *Digital representations of the real world are being used in many applications, such as augmented reality. 6G systems will not only support use cases that rely on virtual worlds but also benefit from their rich contextual information to improve performance and reduce communication overhead. This paper focuses on the simulation of 6G systems that rely on a 3D representation of the environment, as captured by cameras and other sensors. We present new strategies for obtaining paired MIMO channels and multimodal data. We also discuss trade-offs between speed and accuracy when generating channels via ray tracing. We finally provide beam selection simulation results to assess the proposed methodology.*

**Keywords** – 6G, artificial intelligence, machine learning, MIMO, ray tracing.

## 1. INTRODUCTION

Machine Learning (ML) and, more generally, Artificial Intelligence (AI), are currently under investigation to optimize the performance of future communication networks [1]. The applications include, for instance: physical layer (PHY) optimizations, network management and self-organization [2, 3]. Given the increasing importance of ML/AI in communications, there are several initiatives concerning ML/AI architectures, such as the one carried out by ITU [4]. This trend should continue with 6G systems, which are expected to support augmented reality, multisensory communications and high-fidelity holograms [5]. One such application is autonomous driving, where digital representations are used to generate sensors for hardware-in-the-loop testing[1]. And because such digital representations of the world will flow through the 6G network, it is expected that ML/AI can leverage them. Therefore, a specific set of simulation tools for future networks is characterized by the requirement of being able not only of dealing with communication channels, but also the corresponding sensor data, matched to the *scene*.

This paper focuses on strategies for simulating 6G systems that require a representation of the environment, as captured by cameras and, eventually, additional modalities of sensors. More specifically, we consider Multiple Input / Multiple Output (MIMO) systems and discuss the required generation of channels that are consistent with the scene at each time instant. A simulation that integrates communication networks and artificial intelligence immersed in virtual or augmented reality can be computationally expensive, especially for time-varying digital worlds. We discuss two categories of simulations: the one in which the ML/AI model is executed within the virtual world simulation loop and the one in which the ML/AI model is out of the loop and the simulator can then write files to be later used for training ML/AI models. An example of the first category (INLOOP) is going to be used as the *UFPA Problem Statement* [6] for the *2021 ITU AI/ML in 5G Challenge.*

Concerning the channel generation, the requirement of having an associated digital world precludes the adoption of a class of modern channel models that are not related to any virtual world representation, such as the ones presented in [7, 8]. We therefore adopt ray tracing (RT) for MIMO channel generation, which is aligned with other recent work (see, e.g. [1] and references therein) and allows the generation of site-specific communication channel responses with temporal and spatial consistency.

Another motivation for this paper is to promote public datasets. In many ML application domains, the data is abundant or has a relatively low cost. For example, the deep learning-based text-to-speech system presented in [9], which represents the state-of-the-art, achieves quality close to natural human speech after being trained with 24.6 hours of digitized speech. In contrast, the research and development of 5G has to deal with a relatively limited amount of data. Considering the 5G research on AI/ML applied to millimeter waves (mmWave) MIMO, the lack of abundant data from measurements or simulations hinders some data-driven lines of investigation. With 6G moving towards the use of even higher (Terahertz) frequency bands [10], it becomes even more challenging to perform measurement campaigns for this frequency range [11], particularly for outdoor environments. Given that channel measurements for 6G will demand relatively expensive equipment, the simulation strategies for modeling mobility and virtual worlds discussed in this paper can alleviate the problem. The generated datasets are especially useful when spatial consistency and time evolution are important to assess an AI/ML technique applied to the physical layer.

---
[1]https://www.ni.com/pt-br/innovations/white-papers/17/altran-and-ni-demonstrate-adas-hil-with-sensor-fusion.html

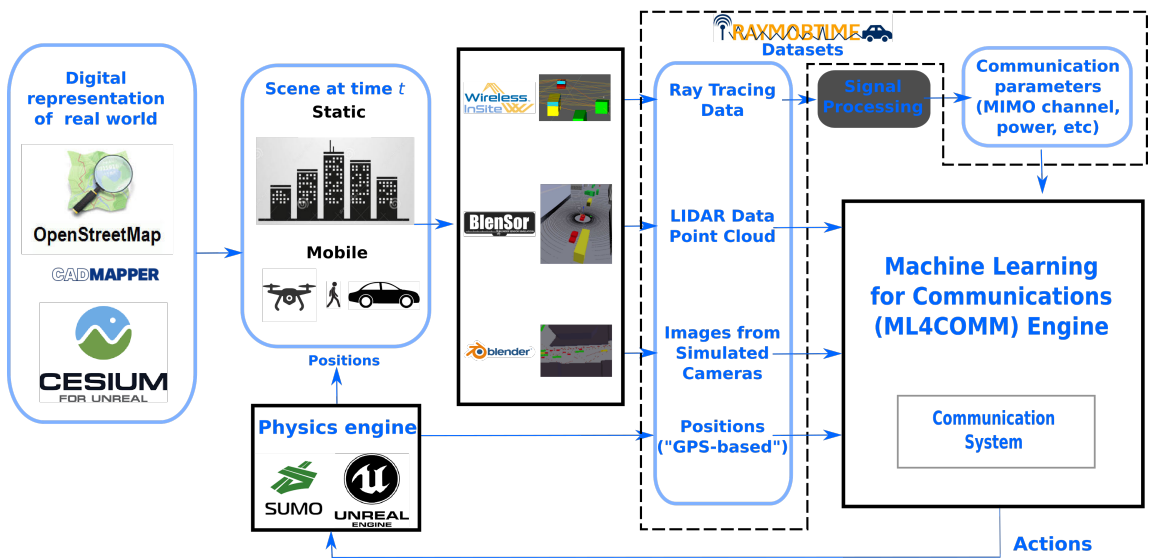

**Fig. 1** – Block diagram of CAVIAR simulation with AI/ML in the simulation loop (INLOOP). In OUTLOOP simulations, the simulator can write files that will be later used for designing and assessing AI/ML models.

The contributions of this paper are:

- A discussion of strategies and software for simulating *Communication networks and Artificial intelligence immersed in VIrtual or Augmented Reality* (CAVIAR).

- A preview of a CAVIAR simulator that will be used in the *UFPA Problem Statement* for the *2021 ITU AI/ML in 5G Challenge*, which consists of a Reinforcement Learning (RL) problem with the decisions taken by the RL agent changing the virtual world on-the-fly (as the simulation evolves).

- Discuss a new methodology using photogrametric data available from the Internet to improve the realism of ray-tracing simulations by automatically assigning electromagnetic properties to the materials composing a scene, via semantic segmentation with deep neural networks.

- Results exposing trade-offs between speed and accuracy when generating channels via ray tracing.

- Results of a reinforcement learning experiment in beam selection realized in the CAVIAR environment.

- Source code and datasets to reproduce the baseline of *2021 ITU AI/ML in 5G Challenge*.[2]

The rest of the paper is organized as follows. Methods and software for CAVIAR simulation of 6G are presented in Section 2. Section 3 explains some improvements in the RT simulation methodology. Section 4 presents numerical results and their discussion. Finally, Section 5 concludes the paper.

## 2. 6G SIMULATION IN VIRTUAL WORLDS

Gaming and other industries are driving the development of sophisticated tools to create virtual worlds, composed of 3D models, physics engines and other components. The virtual world 3D scenery can be created from scratch by 3D design modelers, or from data imported from the real world. For instance, the new *Cesium* plug-in for *Epic Game*'s *Unreal Engine*[3] integrates photogrametric information obtained from drones into 3D models available via *Cadmapper*[4] and other sites. This complements tools such as *Twinmotion*,[5] which facilitate the construction of 3D virtual worlds. This paper promotes the vision that 6G and beyond will benefit from the availability of virtual worlds to leverage ML/AI applied to communication networks. Current investigations of AI applied to 5G aim at finding how raw data from sensors such as LIDAR and cameras can optimize the communication performance [12, 13, 14, 15]. But the possibility of having realistic 3D models, physics engines and other virtual reality assets for simulations of communication systems, opens new horizons in terms of AI/ML applied to 6G and beyond.

---
[2] https://ai5gchallenge.ufpa.br/
[3] https://cesium.com/blog/2021/03/30/cesium-for-unreal-now-available/.
[4] https://cadmapper.com.
[5] https://www.unrealengine.com/en-US/twinmotion.

As proposed in [16], the CAVIAR framework concerns a specific category of 6G simulations that rely on virtual worlds and incorporate two subsystems: wireless communications and AI/ML. In the next paragraphs, we briefly review the CAVIAR framework, depicted in Fig. 1, and then focus on the important aspect of generating the communication channel corresponding to a given scene of the virtual world. We discuss how the Raymobtime methodology [12] fits well to the demand for communication channels imposed by 6G CAVIAR simulations.

A CAVIAR simulation generates multimodal data for each discrete time $t \in \mathbb{Z}$, and is able to operate in two modes, the first mode is focused on online learning, running the simulation and the neural network simultaneously, creating an environment where data is transmitted in real time, or in discrete samples with time stamps defined by the user. The second mode of operation performs data recording in databases or text files, working as a tool for creating datasets. Along the simulation, the *machine learning for communications* (ML4COMM) engine operates on data organized as an *episode* $E = [(\mathcal{P}_1, \mathcal{O}_1), ..., (\mathcal{P}_S, \mathcal{O}_S)]$, with a sequence of S tuples $(\mathcal{P}_t, \mathcal{O}_t), t = 1, ..., S$, of *paired* data, where $\mathcal{P}_t$ and $\mathcal{O}_t$ are sets with the input AI/ML parameters and corresponding outputs, respectively. In supervised learning, $\mathcal{O}_t$ consists of desired labels for classification or regression, while for reinforcement learning $\mathcal{O}_t$ consists of rewards for the agents. The tuples $(\mathcal{P}_t, \mathcal{O}_t)$ denote evolution over discrete-time t. In our methodology, the outputs of the simulators are periodically stored as "snapshots" (or *scenes*) over time $tT_{sam}$, where $T_{sam}$ is the sampling period and $t \in \mathbb{Z}$.

The main steps in Fig. 1 can be summarized as follows. The environment is composed of a 3D scenery with fixed and mobile objects. These objects are created and placed with specialized tools and data from the Internet, as described in [12] and [17]. The positions and interactions among mobile objects are determined by a *physics engine* (for instance, the Unreal engine or the Simulation of Urban MObility (SUMO) traffic generator [18]).

Once the *scene* is complete, the environment is represented via sensors, such as LIDAR, which is simulated by Blensor and Blender software, returning point cloud data (PCD) that maps the shapes of the 3D space around the sensor. It is possible to adjust the resolution of the PCD through a quantization process. A ray-tracing software (Remcom's Wireless InSite in Fig. 1) also captures the communication channel for the given scene. The sensors output constitute the episode input $\mathcal{P}_t$, and the corresponding output $\mathcal{O}_t$ is obtained by a *signal processing* module. These episodes are actually what is stored in Raymobtime episodes [12] but in a CAVIAR simulation they can be created and used on-the-fly, if needed. The CAVIAR 6G virtual world simulator also incorporates a communication system that has some functionalities driven by the ML4COMM engine. The ML4COMM engine also relies on the scene description and can extract features from the raw sensor data to feed its AI/ML algorithms.

Fig. 1 illustrates the INLOOP CAVIAR framework with the AI/ML module within the simulation loop. When the decisions of this module do not affect the environment, it can be convenient to split the simulation into two stages, with the first one being an OUTLOOP CAVIAR simulation that writes *episode* files that will be later used for designing and assessing AI/ML models. The more evolved INLOOP simulation is required in cases such as a drone mission in which the AI/ML decisions will change the drone trajectory and, consequently, its wireless channel. In general, when the AI/ML model issues commands or actuator signals that effectively change the trajectories of mobile entities, alter the environment or the communication system state (e.g., buffer occupation), the simulations may need to be INLOOP and communication channels generated on-the-fly. In the simpler OUTLOOP simulation category, channels can be pre-computed and the communication simulation decoupled from the physical engine, as often used in AI/ML applied to beam selection [19, 12]. The next sections provide two examples to distinguish INLOOP and OUTLOOP CAVIAR simulations.

## 2.1 OUTLOOP CAVIAR simulation for beam selection

Beam selection is a classical application of AI/ML to communications [20, 21, 22]. The goal is to choose the best pair of beams for analog beamforming, with both transmitter (Tx) and receiver (Rx) having antenna arrays with only one Radio Frequency (RF) chain and fixed beam codebooks. Fig. 2 illustrates beamforming from a Base Station (BS) to both vehicles and drones.

We first assume beam selection for a vehicular to infrastructure network, to illustrate an OUTLOOP CAVIAR simulation. In this case the communication subsystem is a downstream MIMO system in which a BS with a Uniform Linear Array (ULA) of $N_t$ antennas communicates with cars with ULAs of $N_r$ antennas. ML is used for beam-selection.

Discrete Fourier Transform (DFT) codebooks $\mathcal{C}_t = \{\bar{\mathbf{w}}_1, \cdots, \bar{\mathbf{w}}_{N_t}\}$ and $\mathcal{C}_r = \{\bar{\mathbf{f}}_1, \cdots, \bar{\mathbf{f}}_{N_r}\}$ are used at the transmitter and the receiver sides, respectively. The beam pair [p, q] is converted into a unique index i ∈ $\{1, 2, \cdots, M\}$, where $M \leq N_t N_r$. For the i-th pair, the *equivalent channel* (without considering noise) can be calculated as

$$y_i = \mathbf{w}_i^* \mathbf{H} \mathbf{f}_i, \quad (1)$$

and the *optimal* beam pair index $\hat{i}$ is given by

$$\hat{i} = \arg \max_{i \in \{1, \cdots, M\}} |y_i|. \quad (2)$$

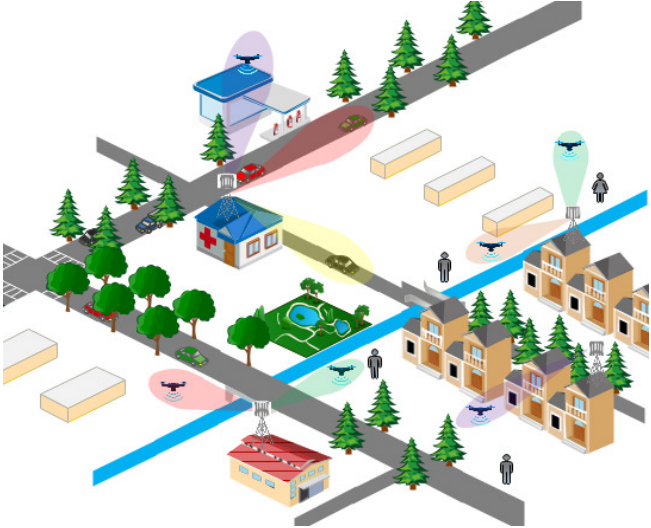

**Fig. 2** – Beamforming from BS to both vehicles and drones.

The beam selection is then posed as a top-k classification problem. At time t, the classifier inputs are features obtained from $\mathcal{P}_t$ and the output is the beam pair i.

For the scenario presented in this section, the trajectory of vehicles and all mobile objects do not depend on the AI/ML model, hence all the episodes can be pre-computed. Next, we discuss a simulation in which the trajectories are determined by the AI/ML model and the channels cannot be pre-computed.

## 2.2 INLOOP CAVIAR simulation with drones and reinforcement learning

Unmanned Aerial Vehicles (UAVs) are being used in many connected applications, such as surveillance and product delivery. UAVs can also be used as mobile radio base stations to extend reach or improve network capacity, mainly in situations of disasters and accidents. In order to meet the requirements of all these use cases, the network links need to obey particular requirements, ranging from very low latency to high data rates [23]. All this motivates intense research on 5G technologies for supporting UAV-based applications. However, there are currently few simulation tools for testing and studying telecommunication systems that involve UAV solutions and their corresponding channels. The CAVIAR framework is deeply integrated with the Unreal Engine development kit and the *Airsim* simulator [24], which bring realism to the physical aspects of the simulations.

As part of the *UFPA Problem Statement* for the *2021 ITU AI/ML in 5G Challenge*, we designed an INLOOP CAVIAR simulation in which RL is executed at the BS and used in two problems: a) determine the drone trajectory and b) beam selection along the downstream. In the challenge, the drones need to deliver pizzas to distinct addresses in a neighborhood. Fig 3 illustrates the scenario.

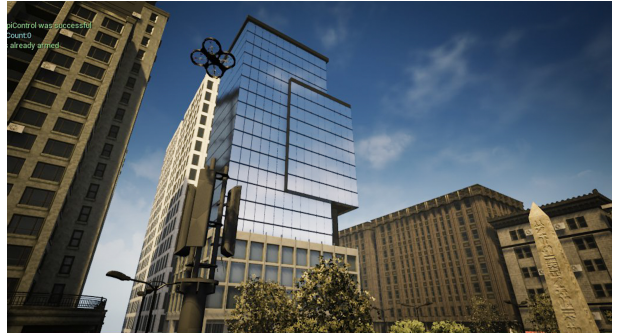

**Fig. 3** – Scene from an INLOOP CAVIAR simulation in which a drone is served by a BS and RL is used for beam selection and for determining the drone trajectory.

The scenario depicted in Fig 3 allows us to investigate several problems that relate communication with drones path planning. One important issue is how to obtain the channels on-the-fly. If the visualization is performed after the whole simulation is finished, the time to generate the channel (via RT, for instance) may be longer. But in this case the scenes need to be visualized along the simulations (as part of a game, for example), then the minimum number of frames per second will impose a limit on the time to generate the communication channels.

The next section discusses our Raymobtime methodology and the corresponding datasets. Other publicly available RT-based datasets are listed in Table 1. The ViWi dataset, presented in [13], provides similar output data compared to Raymobtime, including visual data. The DeepMIMO dataset [25] is maintained by the same group as ViWi and offers only wireless channel information. The dataset described in [1] does not have visual information as well. One of the main differences between these three datasets and Raymobtime is how mobility is handled. The Raymobtime methodology simulates realistic traffic with several moving vehicles using the SUMO software in order to provide better spatial and temporal consistency, as well as channel variability due to the moving scatterers. ViWi [13] (in its first version), DeepMIMO and the map-based channel model in [1] use a fixed grid for Tx-Rx positions and therefore does not consider varying speeds for moving transceivers. ViWi version 2 provides one new scenario that includes several moving vehicles, each equipped with an omnidirectional antenna.

**Table 1** – Other publicly available RT datasets.

| Dataset name | Data Types | Environment | Frequency (GHz) | File format |
|---|---|---|---|---|
| ViWi [13] | Image, depth-map, wireless channel, and user location | Outdoor | 28 and 60 | Matlab, JPEG |
| DeepMIMO [25] | Wireless channel parameters | Indoor and Outdoor | 2.5, 3.5, 28, and 60 | Matlab |
| Map-based channel model [1] | Wireless channel parameters | Indoor and Outdoor | 28 | Matlab |

## 3. IMPROVEMENTS ON RAYMOBTIME METHODOLOGY

The Raymobtime methodology proposed in [12] aims at providing a multimodal dataset, including RT channel information and data from sensors, such as images, LIDAR and location, as illustrated in Fig. 1. One major challenge in building the Raymobtime datasets is to provide accurate wireless communication channel parameter through the use of RT simulation software. In this work, Remcom's Wireless InSite (WI) RT software [26] was adopted given its widespread use [1]. This section summarizes two improvements toward more realistic datasets for AI/ML involving MIMO channels. More details can be found in [16].

The first improvement compared to previous versions of the Raymobtime methodology is the correction of the orientation of the antenna arrays mounted on moving vehicles, so that the array follows the direction of the vehicle. As mobile objects (vehicles, people, etc.) move in the virtual world, previous versions of Raymobtime datasets were not updating the orientation of the antenna array.

The other improvement is the simulation of antenna arrays inside the RT software. Previous versions of Raymobtime always considered omnidirectional antennas inside the RT simulation. This procedure is called here Single Input, Single Output RT (SISO-RT). MIMO channel matrices are obtained during post-processing with the use of the geometrical channel model [27]. This approach reduces processing time and make the dataset more flexible, as the user can define the desired antenna arrays for all transceivers during post-processing, without the need to run RT simulations for every antenna array configuration. However, the geometrical channel model assumes planar-wave propagation, which can be problematic when using large antenna arrays [1]. A more realistic, albeit computationally expensive, alternative is to simulate the antenna arrays inside the RT processing, called MIMO-RT procedure in [16]. Each ray has its own time of arrival and angle offsets, which is equivalent to the spherical-wave assumption [1]. As shown in [28], the difference in estimated MIMO channel capacity can be quite large between the two approaches.

Table 2 presents a list of current Raymobtime datasets and their features. The datasets s011 and s012 include the improvements described in this section. The Raymobtime datasets are divided in several episodes, each one composed by a number of scenes. The smaller the time between scenes, the more similar are consecutive scenes within an episode and, consequently, the more correlated are the communication channels of a given receiver along with the scenes. Currently, RT simulations using Remcom's Wireless InSite (WI) RT software [26] are limited to sub-THz frequencies (up to 100 GHz). More details about the methodology can be found in [12].

The RT simulations demand the identification of the material of the surfaces, in order to properly simulate the electromagnetic interaction of the waves with the objects. The disposition and diversity of these materials directly impact the quality of the channels [29], making this assignment manually a time-consuming and laborious process, and usually results in few materials being actually adopted. To optimize this procedure, the next paragraphs describe ongoing research to develop a methodology to automatically assign such materials to 3D objects via semantic segmentation with deep neural networks.

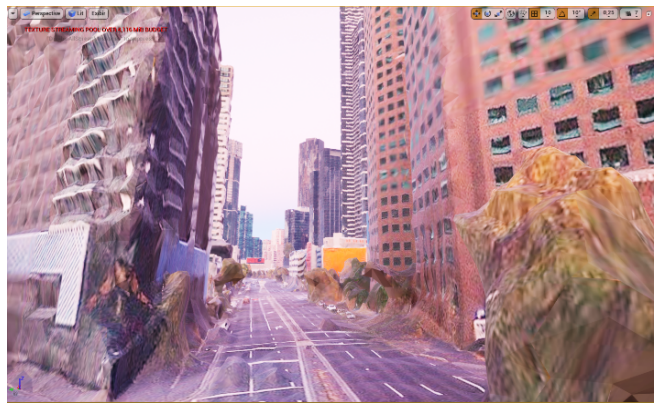

**Fig. 4** – Analysis region image taken from Cesium database.

Semantic segmentation is a modern approach that performs classification at pixel level, and allows us to determine both the class of an object and the boundaries of each object [30]. Current approaches of this method use deep learning in order to overcome traditional object segmentation, allowing us to classify pixels not only by their colors, but also considering the region context [31]. Due to the fact that the 3D environment is built reproducing real locations, it is possible to use databases such as Cesium and Google's Street View to obtain detailed image data from the analysis region. We are applying semantic segmentation in images obtained via the Cesium plug-in for Unreal in order to identify the different

Table 2 – Some Raymobtime datasets.

| Dataset name | Frequency (GHz) | Number of receivers and type | Time between scenes (ms) | Time between episodes (s) | Number of episodes | Number of scenes per episode | Number of valid channels |
|---|---|---|---|---|---|---|---|
| s001 | 60 | 10 Mobile | 100 | 30 | 116 | 50 | 41 K |
| s005 | 2.8 and 5 | 10 Fixed | 10 | 35 | 125 | 80 | 100 K |
| s006 | 28 and 60 | 10 Fixed | 1 | 35 | 200 | 10 | 20 K |
| s008 | 60 | 10 Mobile | - | 30 | 2086 | 1 | 11 K |
| s011 (new) | 60 | 10 Mobile | 500 | 6 | 76 | 20 | 13K |
| s012 (new) | 60 | 10 Fixed | 500 | 6 | 105 | 20 | 21K |

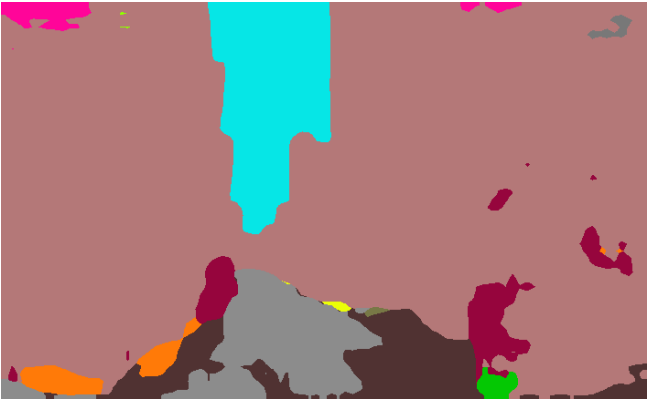

**Fig. 5** – Segmented version using PyTorch of the Cesium image.

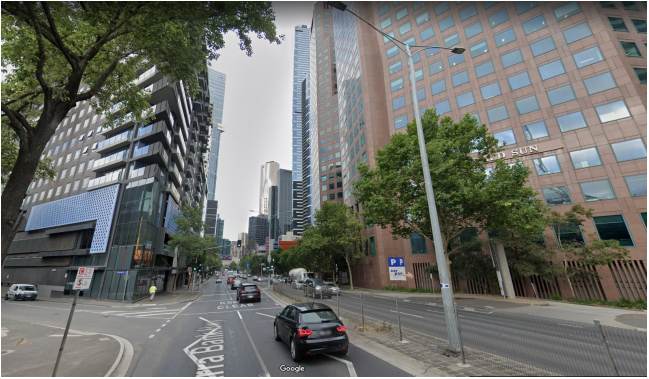

**Fig. 6** – Analysis region image from Google's Street View.

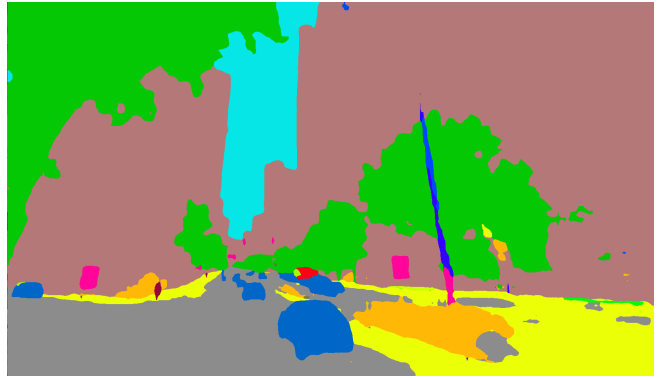

**Fig. 7** – Segmented version of the Google's Street View image.

surface types which composes the scenario.

Fig. 4 and Fig. 5 show an image taken from Cesium and its segmentation, respectively. This segmentation used a PyTorch implementation of semantic segmentation models on the MIT ADE20K [32] scene parsing dataset. In this example, it is possible to verify that the algorithm was capable of determining the contour of the asphalt. On the other hand, the regions corresponding to buildings, cars and vegetation were associated to the same class. This is due to the bad quality of the images taken from Cesium, where some regions of the figure were rendered with deformations and inadequate color assignment to objects, as observed in the tree at the bottom right corner and the objects at the sidewalks, for instance. This is a challenging case for semantic segmentation. In Fig. 6 and Fig. 7, it is possible to verify that there is a significant improvement in the segmentation performance (Fig. 7) compared to the previous example (Fig. 5) when using images obtained from Google's Street View due to the better quality of the source image (Fig. 6). The segmentation was able to identify cars, asphalt, sidewalks, vegetation and buildings with a much better resolution, allowing us to classify the materials with more diversity. Our research efforts are now dedicated to mapping the stitched 2D images to the 3D model and include semantic segmentation results into RT simulations.

## 4. CAVIAR SIMULATION RESULTS

In this section, we discuss some key issues related to CAVIAR simulations. We start by evaluating the computational cost of RT. A snapshot of dataset s012 was simulated with different parameters, assuming isotropic antennas for SISO-RT simulations, and Uniform Linear Array (ULA) for MIMO-RT simulations. The simulations include one transmitter and 10 receivers, each with its own antenna or antenna array, depending on the scenario. The aim is to analyze the impact of the ray spacing, the use of Diffuse Scattering (DS) and the number of antenna elements in the ULA (for MIMO-RT) on the RT simulation time. DS is enabled in all SISO-RT simulations where the carrier frequency is above 6GHz (except for the datasets s011 and s012, as they were designed for the comparison between SISO-RT and MIMO-RT results. The later one has an exponential increase in simulation time when running with DS). For all the simulation results presented here, a PC with an NVIDIA RTX 2070 was used.

In the RT simulations, the transmitter shoots rays in a sphere through the scenario to find viable paths between transmitter and receiver. The minimum angle between the rays is defined as the *ray spacing*. The values in Table 3 show that the ray spacing has a great impact in the total simulation time. For SISO-RT, a simulation using a ray spacing of 0.1° takes 11 times longer than the one with ray spacing equal to 1°. For MIMO-RT, the simulation considering 0.1° ray spacing is 6.2 times longer compared to ray spacing of 1°. For context, Wireless InSite recommends setting ray spacing to 0.2° or less, for 500 m × 500 m areas [33].

DS is a special type of ray interaction with surfaces, allowing for the simulation of scattered paths caused by irregularities in materials. It increases the number of simulated paths and, consequently, the number of calculations and the run time. Table 3 presents results for simulations with DS enabled, both in SISO-RT and MIMO-RT scenarios, considering a ray spacing of 0.5°. For SISO-RT, the run time was 87 longer when enabling the DS compared to not using it. For MIMO-RT this value was even greater: DS increased the simulation time more than 600 times.

As described in Table 4, the simulation times depend on the number of antenna elements in each Tx-Rx pair. Increasing the number of antenna elements in each Tx-Rx pair significantly raises the simulation time. A twelve-fold increase occurs when using $N_t = 64$ and $N_r = 64$ - where $N_t$ and $N_r$ are the number of antenna elements in the ULA of the transmitter and receiver, respectively - compared to the baseline case where $N_t = 64$ and $N_r = 2$.

**Table 3** – Simulation time increase factor for one RT simulation (s012) for different ray spacing values, with and without diffuse scattering enabled. The baseline time for SISO-RT is 00:00:11.749 and for MIMO-RT (with $N_t = 64$ and $N_r = 8$) is 00:00:39.654. The time format is (HH:MM:SS.ccc).

|                    | Simulation time increase factor ||
| ------------------ | ------- | ------- |
| Ray Spacing (°)    | SISO-RT | MIMO-RT |
| 1                  | 1       | 0.7     |
| 0.5                | 1       | 1       |
| 0.25               | 2.4     | 1.5     |
| 0.1                | 11      | 4       |
| 0.5 (DS-enabled)   | 84.7    | 412.9   |

**Table 4** – Simulation time increase factor for one RT simulation (s012) considering different numbers of antenna elements in the transmitter and receiver antenna arrays. The baseline time is 00:00:18.437 (with $N_t = 64$ and $N_r = 2$). The time format is (HH:MM:SS.ccc).

| $N_t$ | $N_r$ | Simulation time increase factor |
| ----- | ----- | ------------------------------- |
| 64    | 2     | 1                               |
| 64    | 8     | 2.2                             |
| 64    | 64    | 12                              |

As an illustration of an INLOOP CAVIAR simulation, we developed code for the Unreal Engine and AirSim to simulate a BS serving a UAV. There are two RL agents: one for determining the UAV trajectory and the other for beam selection. We discuss only the latter agent in this paper. As the UAV flies along its trajectory, the MIMO channel is obtained according to the well-known *geometric model*, with parameters for three multipath components obtained from probability distributions (see, e.g. [27, 16]). This simpler methodology was adopted to speed up the simulations and allow for visualizing the UAV as it flies. In this specific scenario, RT channel responses are not used due to the required simulation time. The BS used a ULA with $N_t = 64$ antennas, while the UAV uses a single antenna. A DFT codebook is adopted.

At each time t, the UAV informs its position to the BS, which can then calculate the Angle of Arrival (AoA) $\theta$ of the beam at the UAV. This angle is used as the input for two beam selection algorithms: one based in RL and a simple baseline. To perform beam-selection using RL, we used a Deep Q Network (DQN) [34]. The Stable Baseline API[6] with default DQN parameters was adopted. The reward is the magnitude of the equivalent channel, as defined in Eq. *(1)*. The baseline algorithm adopts the following heuristic: it simply chooses the beam that points to the straight path direction between the BS and the UAV. For most of the UAV's path, there is Line-Of-Sight (LOS) and this heuristic achieves good results. As expected, this strategy does not work well when the link is Non-LOS (NLOS), which occurs for the angular range $\theta \in [20, 30]$ degrees.

The results of this simple experiment is provided in Fig. 8. The bottom plot shows the angle $\theta$ as the UAV takes off (t $\in [0, 25000]$), reaches its destiny and lands (t > 76000). During three time intervals (including a very short one) the link between the UAV and BS was NLOS. The top plot shows the magnitude of the equivalent channel $|y_{i,t}|$, in which the i-th codebook index was chosen at time t. The optimum value, obtained by exhaustively trying all $N_t = 64$ indices, is shown together with the values obtained by the DQN (RL) and baseline. While the optimum value is always larger than 5 and has an average value of 6.81, both baseline and RL struggle to reach good results and achieve average values $\mathbf{E}[|y_{i,t}|] = 1.7$ and 2.3, respectively. It should be noticed that in this case the RL agent should choose one among 64 indices having a single input (angle $\theta$). In the *UFPA Problem Statement* for the *2021 ITU AI/ML in 5G Challenge* [6], a richer set of input features will be adopted, allowing not only beam selection but also UAV path planning.

---

[6]https://stable-baselines.readthedocs.io/en/master.

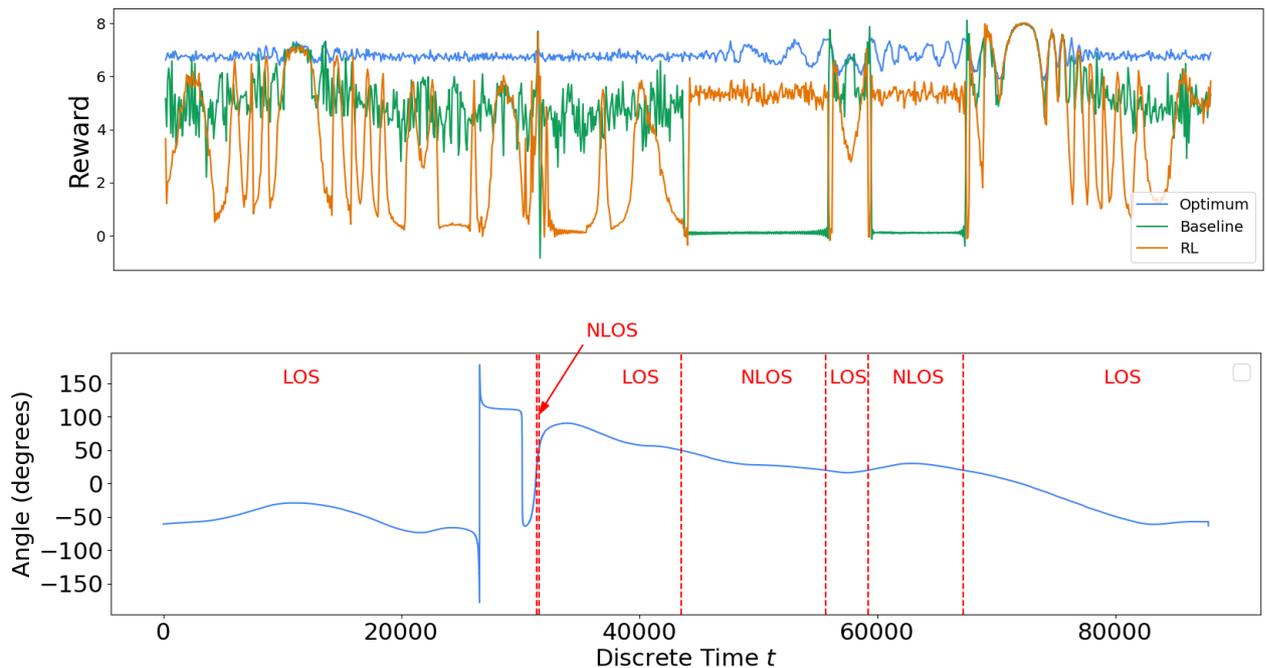

**Fig. 8** – Beam selection results for a BS serving a UAV comparing RL versus a simple baseline algorithm. The optimal result (best beam pair) is also included. The top plot presents the reward is the magnitude of the equivalent channel for the i-th beam pair at the time t (higher reward values are better). The bottom plot shows the AoA $\theta$ at the UAV at each time t.

## 5. CONCLUSIONS

This paper presented strategies and software for simulating 6G systems that represent the surrounding environment with images and other types of data. The so-called CAVIAR framework benefits from virtual reality tools, emphasizing the physical aspects of the movement of objects. This visual information, coupled with MIMO channels generated through RT methods, enables investigating new AI/ML algorithms in 6G that rely on the environment and learning from experience.

We also discussed how semantic segmentation and sensible RT parameters can improve generated MIMO channels. We advocate that aiming at realistic simulations is the natural path to gain a better understanding on how ML/AI can make communication systems more efficient. The effort along the direction of larger and realistic datasets is important for properly evaluating ML-based algorithms, and to avoid unfair comparisons to conventional signal processing.

## REFERENCES

[1] Y.-G. Lim, Y. J. Cho, M. S. Sim, Y. Kim, C.-B. Chae, and R. A. Valenzuela. "Map-Based Millimeter-Wave Channel Models: An Overview, Data for B5G Evaluation and Machine Learning". In: *IEEE Wireless Communications* 27.4 (Aug. 2020). Conference Name: IEEE Wireless Communications, pp. 54–62. ISSN: 1558-0687. DOI: 10.1109/MWC.001.1900315.

[2] Hongji Huang, Song Guo, Guan Gui, Zhen Yang, Jianhua Zhang, Hikmet Sari, and Fumiyuki Adachi. "Deep Learning for Physical-Layer 5G Wireless Techniques: Opportunities, Challenges and Solutions". In: *IEEE Wireless Communications* 27.1 (2020), pp. 214–222. DOI: 10.1109/MWC.2019.1900027.

[3] Chaoyun Zhang, Paul Patras, and Hamed Haddadi. "Deep Learning in Mobile and Wireless Networking: A Survey". In: *IEEE Communications Surveys & Tutorials* 21.3 (2019), pp. 2224–2287. DOI: 10.1109/COMST.2019.2904897.

[4] *Focus Group on Machine Learning for Future Networks including 5G*. https://www.itu.int/en/ITU-T/focusgroups/ml5g/Pages/default.aspx. (Accessed on 03/19/2020).

[5] C. De Lima et al. "Convergent Communication, Sensing and Localization in 6G Systems: An Overview of Technologies, Opportunities and Challenges". In: *IEEE Access* 9 (2021), pp. 26902–26925. DOI: 10.1109/ACCESS.2021.3053486.

[6] LASSE - UFPA. *Radio-Strike: A Reinforcement Learning Game for MIMO Beam Selection*. 2021. URL: https://aiforgood.itu.int/events/radio-strike-a-reinforcement-learning-

[  ] `game-for-mimo-beam-selection-in-unreal-engine-3-d-environments/`.

[7] S. Wu, C. Wang, e M. Aggoune, M. M. Alwakeel, and X. You. "A General 3-D Non-Stationary 5G Wireless Channel Model". In: *IEEE Transactions on Communications* 66.7 (July 2018). Conference Name: IEEE Transactions on Communications, pp. 3065–3078. ISSN: 1558-0857. DOI: `10.1109/TCOMM.2017.2779128`.

[8] J. Bian, C.-X. Wang, X. Gao, X. You, and M. Zhang. "A General 3D Non-Stationary Wireless Channel Model for 5G and Beyond". In: *IEEE Transactions on Wireless Communications* (2021). Conference Name: IEEE Transactions on Wireless Communications, pp. 1–1. ISSN: 1558-2248. DOI: `10.1109/TWC.2020.3047973`.

[9] J. Shen et al. "Natural TTS Synthesis by Conditioning WaveNet on Mel Spectrogram Predictions". In: *https://arxiv.org/abs/1712.05884*. 2018.

[10] Theodore S Rappaport, Yunchou Xing, Ojas Kanhere, Shihao Ju, Arjuna Madanayake, Soumyajit Mandal, Ahmed Alkhateeb, and Georgios C Trichopoulos. "Wireless communications and applications above 100 GHz: Opportunities and challenges for 6G and beyond". In: *IEEE Access* 7 (2019), pp. 78729–78757.

[11] Cheng-Xiang Wang, Jie Huang, Haiming Wang, Xiqi Gao, Xiaohu You, and Yang Hao. "6G wireless channel measurements and models: Trends and challenges". In: *IEEE Vehicular Technology Magazine* 15.4 (2020), pp. 22–32.

[12] A. Klautau, P. Batista, N. Gonzalez-Prelcic, Y. Wang, and R. Heath. "5G MIMO Data for Machine Learning: Application to Beam-Selection using Deep Learning". In: *2018 Information Theory and Applications Workshop (ITA)*. Feb. 2018. URL: `http://ita.ucsd.edu/workshop/18/files/paper/paper_3313.pdf`.

[13] Muhammad Alrabeiah, Andrew Hredzak, Zhenhao Liu, and Ahmed Alkhateeb. "Viwi: A deep learning dataset framework for vision-aided wireless communications". In: *2020 IEEE 91st Vehicular Technology Conference (VTC2020-Spring)*. IEEE. 2020, pp. 1–5.

[14] A. Klautau, N. González-Prelcic, and R. W. Heath. "LIDAR Data for Deep Learning-Based mmWave Beam-Selection". In: *IEEE Wireless Communications Letters* 8.3 (June 2019), pp. 909–912. ISSN: 2162-2345. DOI: `10.1109/LWC.2019.2899571`.

[15] Aldebaro Klautau and Nuria González Prelcic. "Realistic simulations in Raymobtime to design the physical layer of AI-based wireless systems". In: *ITU News MAGAZINE* 5 (2020), pp. 70–73.

[16] Aldebaro Klautau, Ailton de Oliveira, Isabela Pamplona Trindade, and Wesin Alves. "Generating MIMO Channels for 6G Virtual Worlds Using Ray-tracing Simulations (Submitted)". In: *2021 IEEE Statistical Signal Processing Workshop (SSP) (SSP 2021)*. Rio de Janeiro, Brazil, July 2021.

[17] Ailton Oliveira, Marcus Dias, Isabela Trindade, and Aldebaro Klautau. "Ray-Tracing 5G Channels from Scenarios with Mobility Control of Vehicles and Pedestrians". In: *XXXVII Simpósio Brasileiro de Telecomunicações e Processamento De Sinais - SBrT* (2019).

[18] Daniel Krajzewicz, Jakob Erdmann, Michael Behrisch, and Laura Bieker. "Recent Development and Applications of SUMO - Simulation of Urban MObility". In: *International Journal On Advances in Systems and Measurements* 5.3&4 (Dec. 2012), pp. 128–138.

[19] Marcus Dias, Aldebaro Klautau, Nuria González-Prelcic, and Robert W Heath. "Position and LIDAR-aided mmWave beam selection using deep learning". In: *2019 IEEE 20th International Workshop on Signal Processing Advances in Wireless Communications (SPAWC)* (2019), pp. 1–5. DOI: `10.1109/SPAWC.2019.8815569`.

[20] A. Ali, N. González-Prelcic, and R. Heath. "Millimeter Wave Beam-Selection Using Out-of-Band Spatial Information". In: *IEEE Transactions on Wireless Communications* 17.2 (2017), pp. 1038–1052. ISSN: 1536-1276. DOI: `10.1109/TWC.2017.2773532`.

[21] V. Va, J. Choi, T. Shimizu, G. Bansal, and R. W. Heath. "Inverse Multipath Fingerprinting for Millimeter Wave V2I Beam Alignment". In: (2017). IEEE Early access. ISSN: 0018-9545. DOI: `10.1109/TVT.2017.2787627`.

[22] N. González-Prelcic, A. Ali, V. Va, and R. W. Heath. "Millimeter-Wave Communication with Out-of-Band Information". In: 55.12 (Dec. 2017), pp. 140–146. ISSN: 0163-6804. DOI: `10.1109/MCOM.2017.1700207`.

[23] Cheng-Xiang Wang, Ji Bian, Jian Sun, Wensheng Zhang, and Minggao Zhang. "A survey of 5G channel measurements and models". In: *IEEE Communications Surveys & Tutorials* 20.4 (2018), pp. 3142–3168.

[24] Shital Shah, Debadeepta Dey, Chris Lovett, and Ashish Kapoor. "Airsim: High-fidelity visual and physical simulation for autonomous vehicles". In: *Field and service robotics*. Springer. 2018, pp. 621–635.

[25] A. Alkhateeb. "DeepMIMO: A Generic Deep Learning Dataset for Millimeter Wave and Massive MIMO Applications". In: *Proc. of Information Theory and Applications Workshop (ITA)*. San Diego, CA, Feb. 2019, pp. 1–8.

[26] REMCOM. *Wireless InSite*. 2019. URL: https://www.remcom.com/wireless-insite-em-propagation-software.

[27] David Tse and Pramod Viswanath. *Fundamentals of Wireless Communication*. Cambridge University Press, 2005. DOI: 10.1017/CBO9780511807213.

[28] Isabela Trindade, Francisco Müller, and Aldebaro Klautau. "Accuracy Analysis of the Geometrical Approximation of MIMO Channels Using Ray-Tracing". In: *2020 IEEE Latin-American Conference on Communications (LATINCOM)*. IEEE. 2020, pp. 1–5.

[29] Felipe Bastos, Ailton Oliveira, João Borges, and Aldebaro Klautau. "Effects of Environment Model Complexity in Ray-Tracing simulation for UAV Channels". In: *X Conferência Nacional em Comunicações, Redes e Segurança da Informação* (2020).

[30] B. Li, Y. Shi, Z. Qi, and Z. Chen. "A Survey on Semantic Segmentation". In: *2018 IEEE International Conference on Data Mining Workshops (ICDMW)*. 2018, pp. 1233–1240. DOI: 10.1109/ICDMW.2018.00176.

[31] Yanming Guo, Yu Liu, Theodoros Georgiou, and Michael S. Lew. "A review of semantic segmentation using deep neural networks". In: *International Journal of Multimedia Information Retrieval* 7 (2018), pp. 87–93. ISSN: 1536-1276. DOI: https://doi.org/10.1007/s13735-017-0141-z.

[32] MIT Computer Vision team. *ADE20K dataset*. URL: http://groups.csail.mit.edu/vision/datasets/ADE20K/. (accessed: 04.07.2021).

[33] *Wireless InSite Reference Manual*. Version Version 3.3.0. Remcom Inc.

[34] Volodymyr Mnih, Koray Kavukcuoglu, David Silver, Alex Graves, Ioannis Antonoglou, Daan Wierstra, and Martin Riedmiller. "Playing atari with deep reinforcement learning". In: *arXiv preprint arXiv:1312.5602* (2013).

## AUTHORS

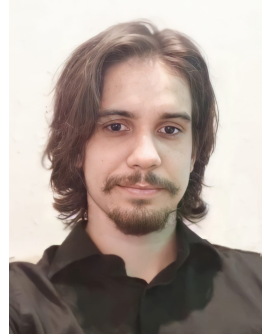

**Ailton Oliveira** is a B.Sc candidate in electrical engineering at Universidade Federal Pará, Brazil. He is currently a research student at the Telecommunications, Automation and Electronics R & D Center (LASSE/UFPA). His achievements were recognized with the outstanding undergraduate researcher award from LASSE/UFPA, and had an awarded article in 2020, by Brazilian Telecommunications Society (SBrT), with a research focused on machine learning applied to beam-selection. His current research interests include digital communications, 5G/B5G networks, MIMO systems, data science and machine learning.

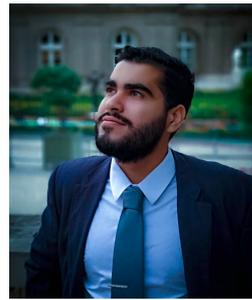

**Felipe Bastos** received a technical degree in telecommunications from Instituto Federal do Pará (2015), Computer Engineer from Universidade Federal do Pará (2020) He also participated in an inter-university exchange at École Supérieure d'Informatique, Électronique, Automatique (ESIEA) through the BRAFITEC program, where he was an intern at the European Nuclear Research Center (CERN). Currently, he is with the Telecommunications, Automation and Electronics R & D Center (LASSE/UFPA) and pursues a Master of Science degree in electrical engineering at Universidade Federal do Pará (UFPA). He is interested in embedded systems, Internet of things, 5G networks, and telecommunication systems.

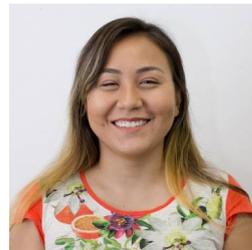

**Isabela Trindade** received a B.E. degree in electrical engineering from Universidade Federal do Pará, Brazil, in 2019. She is currently pursuing a M.Sc. degree in electrical engineering with the Telecommunications, Automation and Electronics R & D Center (LASSE/UFPA), under the supervision of Prof. Aldebaro Klautau. Her research interests include MIMO communications, channel modeling with ray tracing simulations and machine learning.

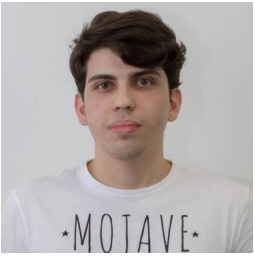 **Walter Frazão** is a B.Sc. candidate at Universidade Federal Pará, Brazil, Brazil. He has been a research student at the LASSE/UFPA since 2019 and a CNPq researcher at the Amazon Center for Excellence in Energy Efficiency (Ceamazon). He received an award in 2020 from the Brazilian Telecommunications Society (SBrT) for an article on machine learning applied to beam selection. His current research interests include 5G, MIMO communications and machine learning.

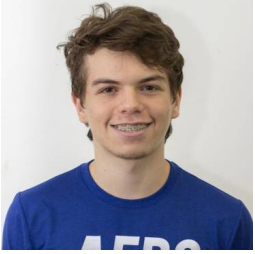 **Arthur Nascimento** started his degree in biomedical engineering in 2018 at the Universidade Federal do Pará. He is currently an undergraduate student researcher at LASSE/UFPA. He received an award for an article in telecommunications in 2020 and also worked on the regional organization of the 2020 ITU AI/ML in 5G Challenge. His current research interests include machine learning, artificial inteligence, computer vision and bioinformatics.

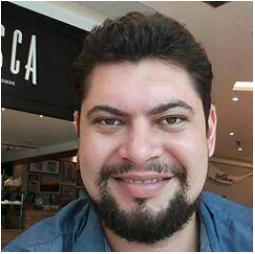 **Diego de Azevedo Gomes** received M.Sc. and Ph.D. degrees in electrical engineering (telecommunications) from Universidade Federal do Pará, Brazil, in 2012 and 2017, respectively. He is currently a professor at Institute of Geosciences and Engineering, Universidade Federal do Sul e Sudeste do Para (Unifesspa). He had two awarded articles in the year 2020, both regarding machine learning applied to beam selection. His current research interests include MIMO communications, digital signal processing, and machine learning.

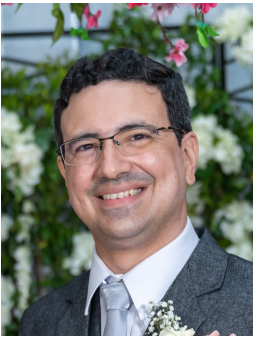 **Francisco Müller** received his Bachelor's degree (2002), M.Sc. (2005) and Ph.D. (2010) in electrical engineering at Universidade Federal do Pará, Brazil. He has been an Associate Professor at Universidade Federal do Pará since 2011 and is associated with the 5G & IoT Research Group at LASSE/UFPA. He was a visiting scholar at Virginia Tech (2007). Current research interests include massive MIMO, 5G/6G channel modeling and estimation. He is a member of the IEEE Communications Society.

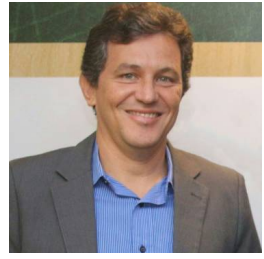 **Aldebaro Klautau** received the Bachelor's (UFPA, Brazil, 1990), M.Sc. (UFSC, Brazil, 1993) and Ph.D. degrees (University of California at San Diego, UCSD, 2003) in electrical engineering. Since 1996, he has been with UFPA and is now a full professor, the ITU-T Focal Point, and directs the 5G & IoT Research Group at LASSE/UFPA. He was a visiting scholar at Stockholm University, UCSD and The University of Texas at Austin. He is a senior member of the IEEE and a researcher of the Brazilian National Council of Scientific and Technological Development (CNPq).